\documentclass[aps,prb,10pt,superscriptaddress,twocolumn,floatfix]{revtex4-1}
\usepackage{amssymb,graphicx,epsfig,amsmath}
\newcommand{\bit}{\begin{itemize}}
\newcommand{\eit}{\end{itemize}}

\newcommand{\SKIP}[1]{} 

\begin{document}

\title{Quantum Monte Carlo Simulations for Stacked Spin Ladder Systems Containing Low Concentrations of Non-Magnetic Impurities; Application to the Low Temperature Broadening of NMR-Spectra in SrCu$_2$O$_3$.}

\author{Robert Achleitner}
\affiliation{Center for Computational Materials Science, Vienna University of Technology, Gu\ss hausstra\ss e 25/134,
A-1040 Vienna, Austria}
\author{Hans Gerd Evertz}
\affiliation{Institute of Theoretical and Computational Physics,
Graz University of Technology, Petersgasse 16, 8010 Graz, Austria}
\author{Masatoshi Imada}
\affiliation{Department of Applied Physics, University of Tokyo,
Hongo, Bunkyo-ku, Tokyo, 113-8656, Japan}
\author{Ralf Gamillscheg}
\affiliation{Institute of Theoretical and Computational Physics, Graz University of Technology, Petersgasse 16, 8010 Graz, Austria} 
\author{Peter Mohn}
\affiliation{Center for Computational Materials Science, Vienna University of Technology, Gu\ss hausstra\ss e 25/134,
A-1040 Vienna, Austria}
\date{\today}

\begin{abstract}
We present a quantum Monte Carlo study for Heisenberg spin-$\frac{1}{2}$ two-leg ladder systems doped with non magnetic impurities. The simulations are applied to the doped spin-ladder compound Sr(Cu$_{1-x}$Zn$_x$)$_2$O$_3$ where a large broadening of the $^{65}$Cu NMR lines has been observed in experiment
at low temperatures far larger than the N\'eel temperature.
We find that interladder couplings with a sizable coupling in the stacking direction are required to describe the line broadening, which can not be explained by considering a single ladder only. 
Around a single impurity, spin correlations cause an exponentially decaying antiferromagnetic local magnetization in a magnetic field.
We develop an effective model for the local magnetization of systems with many randomly distributed impurities, with few parameters which can be extracted out of quantum Monte Carlo calculations with a single impurity.
The broadening arises from a drag effect, where the magnetization around an impurity works as an effective field for spins on the neighboring ladders,
causing a non-exponentially decaying magnetization cloud around the impurity. 
Our results show that even for impurity concentrations as small as $x=0.001$ and $x=0.0025$, the broadening effect is large, in good quantitative agreement with experiment.
We also develop a simple model for the effective interaction of two impurity spins.
\end{abstract}

\maketitle

%%%%%%%%%%%%%%%%%%%%%%%%%%%%%%%%%%%%%%%%%%%%
\section{Introduction}
%%%%%%%%%%%%%%%%%%%%%%%%%%%%%%%%%%%%%%%%%%%%
%

SrCu$_2$O$_3$ is a spin-$\frac{1}{2}$ Heisenberg spin-ladder compound that has been studied intensely both experimentally and theoretically. The crystallographically determined structure\cite{azuma2004pre,SrCu2O3structure2006} 
consists of planar Cu-O trellis lattices with intercalated Sr ions. These trellis lattices contain the spin ladders which are almost perfectly decoupled due to frustration. The unpaired electron of the Cu$^{2+}$ ion carries a spin-$1/2$ and the spin dynamics arises from the Cu-O-Cu interaction via superexchange over the oxygen ions. The hyperfine interaction couples the magnetic moments of the spins to the nuclear magnetic moments of the Cu ions and in turn influences the local magnetic resonance field. Upon doping with very
small amounts of non-magnetic impurities, e.g. 0.25\% of Zn, which occupy Cu sites, a 
surprisingly large broadening of the Cu NMR-spectrum with decreasing temperature has been observed.\cite{PhysRevLett.80.604, PhysRevB.60.4181} 
A satisfactory theoretical explanation is still missing.
It is known that an impurity in a single spin ladder causes an exponentially decaying staggered effective local magnetic moment profile around this impurity.\cite{IinoImada96,JPSJ.65.1949,JPSJ.65.2377,Mikeska97,Sandvik97,PhysRevB.57.10755,Greven98,Bobroff09,RevModPhys.81.45} However, fitting the NMR spectra on the basis of this exponential behavior requires much larger correlation lengths 
than found in theoretical studies of single spin ladders.\cite{PhysRevLett.77.1865,Greven98} 
For a reasonable fit with an impurity concentration of $x = 0.0025$, Ref.~\onlinecite{PhysRevLett.80.604} suggested a correlation length of about $\xi_x$ $\sim$ 100,
and for a concentration of $x = 0.001 - 0.003$, Ref.~\onlinecite{PhysRevB.60.4181} estimated  $\xi \sim$ 20 - 50.
Theoretical studies suggest, however, that the correlation length of undoped ladders is much smaller,
and they show that it barely changes upon introduction of a few impurities.\cite{Sandvik97,Mikeska97,Greven98,Bobroff09} 
For $Bi Cu_2 PO_6$, a differently structured material with large inter-ladder coupling inside one layer, an exponentially decaying cloud model\cite{Bobroff09,PhysRevB.81.054438} 
($\sim$exp(-($\xi_x$/$r_x+\xi_y$/$r_y$)) and 
a stacked ladder version ($\sim$exp(-($\xi_x$/$r_x+\xi_y$/$r_y+\xi_z$/$r_z$)) 
are reported to show qualitative agreement with experiment
for impurity concentrations around $x = 0.02$, 
but fail to explain the broadening at very small dilutions ($x \leq 0.005$). 

In the present paper we perform quantum Monte Carlo (QMC) simulations for single and stacked spin-ladders within the parameter range suggested in the report by Johnston \textit{et al}.\cite{johnston-2000}. 
We develop an effective model for the magnetization on systems of stacked ladders with
random impurities which needs only a few parameters measured by QMC
and permits an efficient calculation of the NMR spectrum.
We find that the coupling of adjacent stacked spin ladders 
strongly influences the NMR spectrum, and a value at the upper end of
the range suggested in Ref.~\onlinecite{johnston-2000}
is required to describe the experimentally found low-temperature NMR line broadening down to very small impurity concentrations, 
consistent with a study on the chain material Sr$_2$CuO$_3$.\cite{SirkerLaflorencie09}

In Sec. II, we briefly summarize the results of NMR experiments on $SrCu_2O_3$ with nonmagnetic impurities. 
In Sec. III, we specify the Heisenberg model which we use to describe this material.
Section IV contains results on single ladders, including an effective model for the
interaction of two impurities,
and Sec. V contains our results on stacked ladders.
Sec. VI discusses the effects on NMR spectra, and Sec. VII contains our conclusions.

%%%%%%%%%%%%%%%%%%%%%%%%%%%%%%%%%%%%%%%%%%%%
\section{NMR experiment}\label{experiment}
%%%%%%%%%%%%%%%%%%%%%%%%%%%%%%%%%%%%%%%%%%%%
%
%"""""""""""""""""""""""""""""""""""""""""""""""""""""""""""""""""""""''
\begin{figure}[b]
   \centering
	\includegraphics*[width=0.48\textwidth]{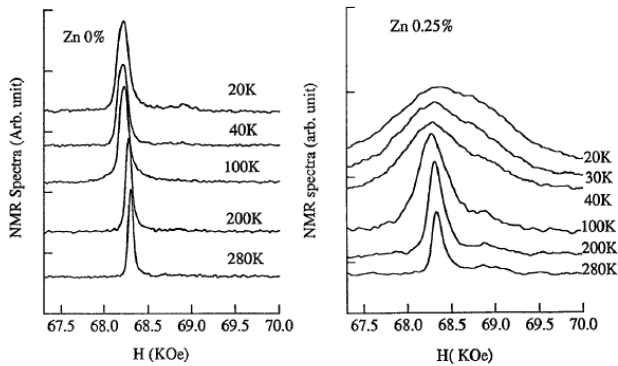}
	\caption{Experimental NMR spectra from Fujiwara et al.\cite{PhysRevLett.80.604} The right panel shows the NMR linewidth broadening in doped SrCu$_2$O$_3$, while for the undoped sample (left panel) this effect is almost absent.}
	\label{fig:ExpFuji} 
\end{figure}
%"""""""""""""""""""""""""""""""""""""""""""""""""""""""""""""""""""""''
%
In their experimental investigation, Fujiwara et al. \cite{PhysRevLett.80.604} reported NMR studies on undoped and doped SrCu$_2$O$_3$. The results are shown in Fig.~\ref{fig:ExpFuji}. For a doping of $x=0.0025$ a massive broadening of the Cu NMR spectrum is observed upon lowering the temperature, while in the undoped specimen this effect is almost absent. 
Similar experimental results have been obtained for SrCu$_2$O$_3$ by Ohsugi \textit{et al.}, \cite{PhysRevB.60.4181}
and also for other spin-ladder systems YBa$_2$(Cu$_{1-x}$Zn$_x$)$_3$O$_{6+y}$,\cite{PhysRevLett.67.3140,JPSJ.62.2803} YBa$_2$(Cu$_{1-x}$Zn$_x$)$_4$O$_{8}$, \cite{PhysRevB.61.4319} and Bi(Cu$_{1-x}$Zn$_x$)$_2$PO$_6$ .\cite{PhysRevLett.105.067203} 

NMR measures the distribution of magnetizations in a system exposed to an external magnetic field.
We note that quantitatively, the broadening of the NMR signal corresponds to %is determined by 
relatively {\em small} magnetizations 
on the order of $0.001\dots 0.01\,\mu_B$ (see Sec. \ref{secProfile2NMR}).
At first sight an increasing linewidth for decreasing temperature is counterintuitive. 
It suggests that the observed broadening is due to the temperature dependence of correlations along and between the spin ladders,
and Refs~\onlinecite{PhysRevLett.80.604,PhysRevB.60.4181} phenomenologically described 
the broadening with very large spin correlation lengths.
However, the broadening takes place at temperatures far above the 3d ordering temperature of about
1 K in SrCu$_2$O$_3$ at the same impurity concentration,\cite{Azuma97,larkin_crossover_2000}
where long-range correlations would be expected.

In the following we will therefore investigate the distribution of magnetizations in a model for  SrCu$_2$O$_3$ with impurities.
We find that the spin correlation length is indeed barely affected by impurities in the temperature range of the
NMR experiments, but the magnetization profiles around impurities is strongly influenced by the coupling of stacked ladders.

%%%%%%%%%%%%%%%%%%%%%%%%%%%%%%%%%%%%%%%%%%%%
\section{Modeling S\MakeLowercase{r}C\MakeLowercase{u}$_2$O$_3$} \label{modelling}
\begin{figure}[b] 
    \centering
	\includegraphics*[width=0.45\textwidth]{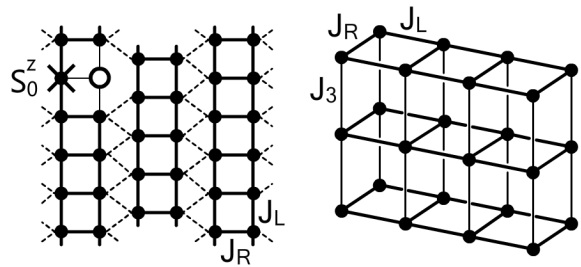}
	\caption{Model structure for SrCu$_2$O$_3$. Left panel: Trellis lattice in the Cu-O plane. 
                The effective ladders (solid lines) are decoupled by frustration, symbolized by the dashed lines. 
                We denote by S$_0$ (cross) a spin located on the same rung as a non-magnetic impurity (open circle). 
                Right panel: Stacked ladders coupled via J$_{3}$.}
	\label{fig:TrellisLat}   
\end{figure}

%%%%%%%%%%%%%%%%%%%%%%%%%%%%%%%%%%%%%%%%%%%%
%
The unsaturated spins of the Cu$^{2+}$ ions in the Cu-O planes in SrCu$_2$O$_3$ form spin ladders which are visualized in Fig.~\ref{fig:TrellisLat}, where the spins along the ladder couple anti-ferromagnetically. 
Within the Cu-O planes the ladders form a trellis lattice,\cite{PhysRevB.49.8901} 
while normal to the Cu-O planes the ladders are stacked. The trellis lattice is responsible for an effective decoupling of the ladders due to frustration. 
We model SrCu$_2$O$_3$ like in Refs.~\cite{PhysRevB.54.13009,johnston-2000} 
as a system of stacked ladders with spin-$\frac{1}{2}$ Heisenberg interactions,

\begin{equation}
 \hat{\cal H} = k_B \sum_{ij} J_{ij}\, \vec S_i \vec S_j - \mu_B g H \sum_i S_i^z \;,
\end{equation}
with nearest-neighbor couplings $J_{ij}$ and an external magnetic field $H$,
\footnote{In specifying $H/J_L$ in the paper we leave out the unit Tesla/Kelvin, 
i.e. $H/J_L=0.01$ should be read as $(\mu_B H)/(k_B J_L) = 0.672 * 0.01$.}
$k_B\simeq 1.38\, 10^{-23}J/K$, $\mu_B \simeq 9.27\, 10^{-24} J/T$, and $g=2$.

Inside the ladder the interaction is described by the coupling $J_L$ along the ladder leg  and $J_R$ along the rungs. The difference in the electronic structure of the oxygen ions on the rungs and on the ladder legs causes an anisotropy of the spin coupling constants J$_{L}$ and J$_{R}$\cite{PhysRevB.58.R14713}. In the stacking direction the interaction is described by 
a coupling J$_{3}$. 
We denote the $x$ direction to be along the ladder and the $z$ direction along the stacking direction. 
Doping by a Zn atom introduces a non-magnetic impurity, i.e., a missing site in the model.

We simulate the model using a highly parallelized \cite{web:OpenMPI} Quantum Monte Carlo code with a directed loop algorithm \cite{PhysRevLett.58.86,evertz03,PhysRevLett.70.875,PhysRevE.66.046701} 
in stochastic series expansion (SSE) representation of the associated path integral.
We employ the spin ladder structure as outlined in Fig.~\ref{fig:TrellisLat},
periodic boundary conditions in the chain (=x) direction, 
and for stacked ladders also in the stacking (=z) direction.
%
%"""""""""""""""""""""""""""""""""""""""""""""""""""""""""""""""""""""''
%"""""""""""""""""""""""""""""""""""""""""""""""""""""""""""""""""""""''
%

We follow the results of Johnston at al.\cite{johnston-2000} for the size of interactions.
\footnote{Potential additional cyclic couplings \cite{Calzado03} are not accessible by QMC calculations because of a severe sign problem.}
For a single ladder we use $J_R$/$J_L=$ 0.4 - 0.6 with $J_L=$ 1905 K. 
For stacked ladders,  we employ $J_R$/$J_L=$ 0.5 and $J_3$/$J_L=$ 0.001 - 0.03 with $J_L=$ 1920 K. 
The magnetic field used in the NMR experiments is O(10) T; i.e.,\ $H/J_L \simeq 0.003$.
%

%
%"""""""""""""""""""""""""""""""""""""""""""""""""""""""""""""""""""""''
\begin{figure}[t]      
\centering
	\includegraphics*[width=0.45\textwidth]{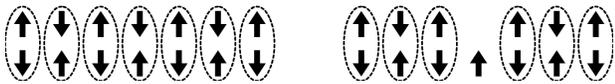}
	\caption{
Rung picture of a spin ladder without (left) and with (right) an impurity. Spins form singlets with a spin gap which prevents the system from responding to a small external field. An impurity breaks one singlet and leaves a free spin-$1/2$ which responds to small external fields and couples to surrounding spins.}%caption	
\label{fig:RungPic}
\end{figure}
%"""""""""""""""""""""""""""""""""""""""""""""""""""""""""""""""""""""''

%%%%%%%%%%%%%%%%%%%%%%%%%%%%%%%%%%%%%%%%%%%%
\section{Single ladder}  \label{subsec:singlelad}
%%%%%%%%%%%%%%%%%%%%%%%%%%%%%%%%%%%%%%%%%%%%

%-------------------------------------------------------
\subsection{Magnetization profile of a single impurity}
%-------------------------------------------------------
%
%

%=======================================================
To describe the impact of an impurity in a spin ladder one may first consider the case of $J_R\gg J_L$ (or vanishing J$_L$) which leads to the so called rung picture,\cite{DagottoRice96} 
in which the rungs are seen as independent from each other (Fig.~\ref{fig:RungPic}). The spins of a rung form a singlet separated by a spin-gap of energy $J_R/2$ from the triplet state that prevents the system from responding to a small external magnetic field. By introducing an impurity, one singlet is broken up and leaves a spin-$1/2$ free to respond to an external magnetic field (free spin). The picture remains useful even at large $J_L$, where the presence of an impurity still breaks a singlet and leaves a free spin.\cite{RevModPhys.81.45}
%=======================================================

\begin{figure}[t]      
\centering
\includegraphics*[width=0.45\textwidth]{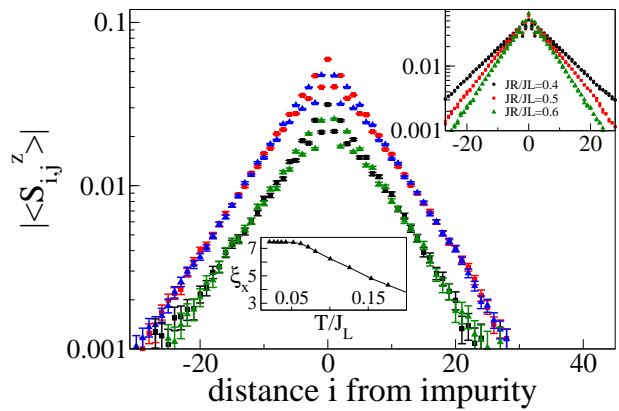}
	\caption{
Local spin expectation value (absolute value) 
of a single ladder for $J_R/J_L=0.5$, $H = 0.01 J_L$ %($\simeq 30$ Tesla). 
with a single impurity, at two different temperatures. 
Upper curve (red/blue) $T = 0.02 J_L \simeq 40$ K, 
lower curve (black/green) $T = 0.05 J_L \simeq 100$ K. The different colours refer to magnetic moments located on the undoped ladder leg (red and black circles) and on the doped ladder leg (blue and green triangles), respectively. For the two temperatures considered, the  correlation length remains almost constant (lower inset). 
The upper inset 
shows the spin expectation values at $T/J_L=0.02$ % 0.02083
for different values $J_R/J_L$. 
The correlation length $\xi_x(J_R/J_L)$ becomes $5.9$, $7.45$, and $9.75$ for $J_R/J_L=$ $0.6$, $0.5$, and $0.4$, 
respectively.}%caption	
\label{fig:LadderLegsOverLay}
\end{figure}

In a magnetic field, the antiferromagnetic correlations on the ladder cause a staggered local magnetic moment profile to develop around the impurity. 
In Fig.~\ref{fig:LadderLegsOverLay} the absolute values of the local spin expectation values are plotted for two temperatures
within the experimental range (see Fig.~\ref{fig:ExpFuji}). 
The profiles have their maximum at the spin S$_0$ residing on the same rung as the impurity and drop exponentially with the correlation length $\xi_x$ of the undoped system.\cite{Sandvik97,Mikeska97,Greven98,Bobroff09} 
as
\begin{equation}\label{eq:expdecay}
	<S_{i,j}^z>=<S_{0}^z> (-1)^{i+j} e^{-\frac{\vert i \vert}{\xi_x}} \quad .
\end{equation}
Here, $S_{0}^z=S_{0,0}^z$ is the spin on the same rung as the impurity and $i$ is the distance in the leg direction and $j=\lbrace 0,1\rbrace$ in the rung direction 
from S$_0$.
Plotting $\vert \langle S_{i,j}^z \rangle \vert$ on a logarithmic scale (Fig.~\ref{fig:LadderLegsOverLay}) shows the nearly perfect exponential dependence.
Some deviations occur close to the impurity, at large magnetizations not directly relevant for the observed broadening 
 (see Sec. \ref{secProfile2NMR}).
The absolute value of the spins on the same ladder leg as the impurity ($j=1$) is somewhat smaller
than on the other leg at small distance $i$, 
but it appears to approach the values of the $j=0$ leg at large distances $i$.

The lower inset of Fig.~\ref{fig:LadderLegsOverLay} shows the temperature dependence of the correlation length.
It remains almost constant \cite{PhysRevLett.77.1865,Greven98,Bobroff09} below $T\simeq 0.05 J_L \simeq 100K$.
The upper inset in Fig.~\ref{fig:LadderLegsOverLay} compares the profiles for varying $J_R/J_L$. Upon decreasing $J_R/J_L$, the correlation length increases, however even at $J_R/J_L=0.4$ the resulting correlation length is much smaller than the values fitted to the NMR spectra in earlier studies.\cite{PhysRevLett.80.604,PhysRevB.60.4181}
Correspondingly, the exponentially decaying clouds around impurities on independent ladders
shown in Fig.~\ref{fig:LadderLegsOverLay}
produce only a very small broadening of the NMR signal (see Sec. \ref{secProfile2NMR}).

%"""""""""""""""""""""""""""""""""""""""""""""""""""""""""""""""""""""''
%"""""""""""""""""""""""""""""""""""""""""""""""""""""""""""""""""""""''
%

%-------------------------------------------------------
%\subsection{Temperature dependence} \label{sec:tempbehave}
\subsection{Temperature and magnetic field dependence} \label{sec:tempbehave}
\begin{figure}[t]
    \centering
%	\vspace{4mm}
	\includegraphics*[width=0.45\textwidth]{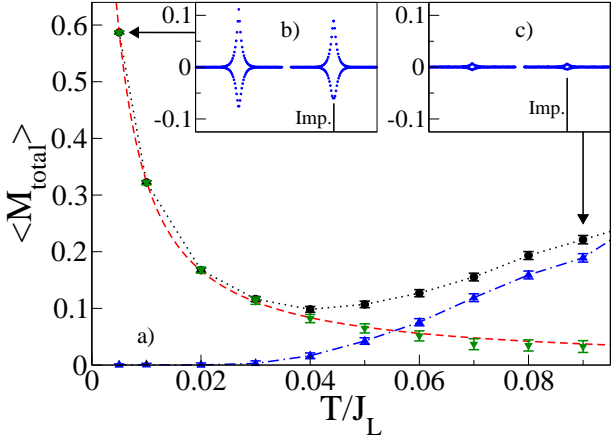}
	\caption{
 (a) Total magnetization versus $T/J_L$ for a system in a magnetic field, 
      with (blue up-triangles) and without (black circles) an impurity. 
    System parameters: 200 $\times$ 2 ladder,  $J_R$/$J_L=$ 0.5, $H = 0.005 J_L$ ($\simeq 14.3$ T). 
    To observe the effect caused by the impurity only, the difference between the undoped and doped systems 
    is also shown (green down-triangles). The red dashed line is the analytical solution for a free spin in a magnetic field.
     Blue and black lines are guides to the eye.
   In the insets, the staggered distributions of local magnetic moments around the impurity
   at (b) $T= 0.005 J_L \simeq 9.6$ K and (c) $T = 0.09 J_L\simeq 173$ K are plotted.
   (For each inset: ladder leg with impurity on the right, leg opposite impurity on the left.)
}
	\label{fig:MvsT}
\end{figure}

%-------------------------------------------------------
%
We analyze the temperature dependence of spin magnetizations \cite{Sandvik97} 
in systems with a single impurity and without impurities.
We observe two mechanisms which have a direct impact on the NMR spectrum. 
In Fig.~\ref{fig:MvsT}, 
both effects are demonstrated for the total magnetization 
            $M_\text{total} = g \sum_{ij} \langle S^z_{ij} \rangle$
of the doped and undoped systems. 
% (1)
The undoped system (blue up-triangles) exhibits a collective excitation of all the spins as a response to the external field 
at temperatures above about $0.04 J_L \simeq 80K$ (blue line in Fig.~\ref{fig:MvsT}).
This collective excitation causes a temperature dependent shift of the NMR peak without changing its  width,
in quantitative agreement with the NMR results in Fig.~\ref{fig:ExpFuji}.      
% (2)
The second effect is the creation of a local cloud of magnetic moments around the impurity
[Figs.~\ref{fig:LadderLegsOverLay} and \ref{fig:MvsT}(insets)]
which causes the total magnetization to grow again at sufficiently low temperatures,
corresponding to a shift back of the NMR signal in  Fig.~\ref{fig:ExpFuji} 
at temperatures of 40 K ($0.02 J_L$) and below.
By plotting the difference of the total magnetization between the undoped and the doped systems 
%(Fig.~\ref{fig:MvsT} green down-triangles) 
we find that the total magnetization caused by the impurity 
corresponds to one spin $\frac{1}{2}$ moment \cite{IinoImada96,Miyazaki97,Sandvik97}
at low temperature, in agreement with experiment,\cite{Azuma97} and that it closely follows the analytic solution of a free spin in a magnetic field
(dashed red line in Fig.~\ref{fig:MvsT}) 
$M_\text{total} \simeq g \langle S^z_{free} \rangle$, with
\begin{equation}
  \langle S^z_{free} \rangle = \frac{1}{2} \tanh (\frac{\mu_B H}{k_B T}) \quad ,
  \label{equ:freespin}
\end{equation}
independent of the size of couplings, in the range studied.

%"""""""""""""""""""""""""""""""""""""""""""""""""""""""""""""""""""""''
%"""""""""""""""""""""""""""""""""""""""""""""""""""""""""""""""""""""''
%

%"""""""""""""""""""""""""""""""""""""""""""""""""""""""""""""""""""""''
\begin{figure}[t]
%   \vspace{4mm}
   \centering
	\includegraphics*[width=0.47\textwidth]{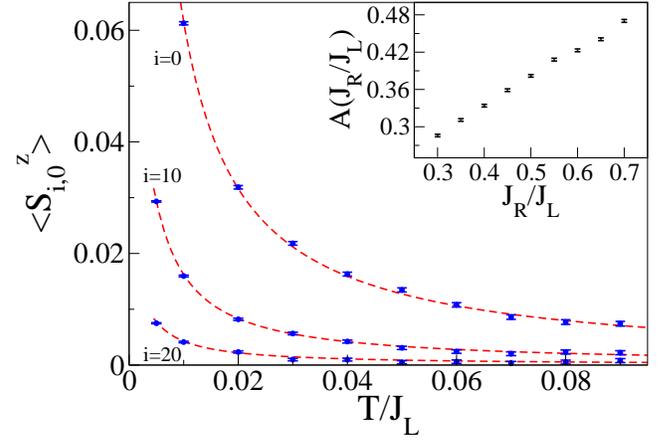}
	\caption{Spin expectation values of the spins opposite of the impurity (i=0) and at a distance of
     10 lattice sites (i=10) and 20 lattice sites (i=20) versus T/${J_L}$, all on the ladder leg opposite the impurity. 
     The dashed lines correspond to a free spin in  a magnetic field times a proportionality factor. 
     System parameters: 200 $\times$ 2 ladder, $J_R$/$J_L=$ 0.5, $H$/$J_L=$ 0.005. 
     The inset shows the factor $A(\xi_x)$, measured at $T=0.02 J_L$.}
	\label{fig:S0vsT}
\end{figure}
%"""""""""""""""""""""""""""""""""""""""""""""""""""""""""""""""""""""''
%

%

Indeed, we find that the individual local magnetic moments also show a \textit{tanh} behavior, 
modulated by the exponential decay of Eq.~\ref{eq:expdecay},
\begin{equation}
  \langle S^z_{i,j}(T) \rangle = A(\xi_x) (-1)^{i+j} e^{-\frac{\vert i  \vert}{\xi_x}} \frac{1}{2} \tanh (\frac{\mu_B H}{k_B T})
  \label{equ:stag_w_corr}
\end{equation}
with a proportionality factor $A(\xi_x)$ which depends on the correlation length.

In the temperature range $T<0.05 J_L \simeq 100K$ where most of the broadening takes place in experiment,
$\xi_x$ is independent of temperature (Fig.~\ref{fig:LadderLegsOverLay}, lower inset).
In this range, $A(\xi_x)$ can be obtained from a single QMC simulation by measuring $\langle S_0^z\rangle$
%i.e.\ $i=0$ and $j=0$ 
at a reference temperature $T_{ref}$ and a magnetic field H$_{ref}$: 
%from Eq.~\ref{equ_A_of_xi}
%
\begin{equation}
A(\xi_x) = \frac{\langle S^z_{0}(T_{ref}/J_L) \rangle}{\frac{1}{2} \tanh (\frac{\mu_B H_{ref}}{k_B T_{ref}})} \quad .
\label{equ_A_of_xi}
\end{equation}

%"""""""""""""""""""""""""""""""""""""""""""""""""""""""""""""""""""""''
\begin{figure}[t]
    \centering
%	\vspace{4mm}
	\includegraphics*[width=0.45\textwidth]{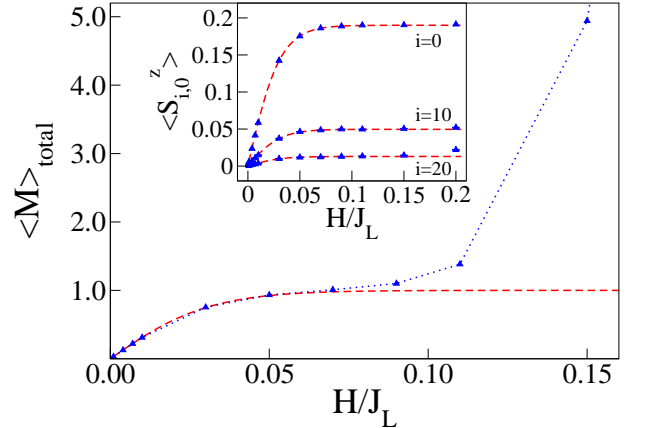}
	\caption{Total magnetization versus H/$J_L$ for a system with a single impurity (blue up-triangles). System parameters: 300 $\times$ 2 ladder,
 $J_R$/$J_L=$ 0.5, $T_{ref}$/$J_L=$ 0.02083. The red dashed line corresponds to the solution of a free spin in a magnetic field. The inset compares Eq.~\ref{equ:stag_w_corr} with the QMC results for $\langle S_{i,0}^z \rangle$,
        with $A(\xi)$ measured at $H_{ref}=0.01 J_L$ and the same $T_{ref}$.}
	\label{figMvsH}
\end{figure}
%"""""""""""""""""""""""""""""""""""""""""""""""""""""""""""""""""""""''

%------
% Figure 6
%------
Fig.~\ref{fig:S0vsT} shows a comparison between the analytical model Eq.~\ref{equ:stag_w_corr} and QMC results for three different lattice sites on the ladder leg opposite to the impurity (j=0).
$A(\xi_x)$ was extracted from a single QMC simulation at $T_{ref}= 0.005 J_L \simeq 10$K. 
In the broadening temperature regime ($\xi_x \sim $ const.) and even beyond, the analytical model (dashed lines) matches the QMC data very well.

%This also holds for stacked ladders if the correct magnetic moment profile is applied (see Sec. %\ref{sec1ImpStackedDescription}).

%For systems where $\xi_x$ shows a pronounced temperature dependence the relation given in Eq.~\ref{equ:stag_w_corr} 
% is only valid %for $\xi_x \rightarrow \xi_x(T)$ and thus $A(\xi_x) \rightarrow A(\xi_x(T))$. 
%In this case QMC simulations are required for each %temperature step. 

On the ladder leg containing the impurity (j=1), the magnitude of local spin expectation values is somewhat smaller than
specified in Eq.~\ref{equ:stag_w_corr} (see Fig.~\ref{fig:LadderLegsOverLay}),
consistent with the total homogeneous magnetization of the ladder to be that of a single free spin without a factor $A$.
[Eq.~\ref{equ:freespin} and Fig.~\ref{fig:MvsT}].

%
%-------------------------------------------------------
%\subsection{Magnetic field dependence}
%-------------------------------------------------------
%
%
%We perform calculations of a ladder with a single impurity for different magnetic fields H/J$_L$ and fixed temperature. 
In Fig.~\ref{figMvsH} we compare Eqs.~\ref{equ:freespin} and \ref{equ:stag_w_corr} 
to QMC results as a function of magnetic field $H$ at fixed temperature.
%
%(blue triangles) for the total magnetic moment of the ladder to the free spin solution Eq.~\ref{equ:freespin} (red %dashed line). 
%The inset shows the QMC results for the local spins (blue up-triangles) opposite of the impurity (i=0) 
%and at distant sites i=10 and i=20 in comparison to the solution given by Eq.~\ref{equ:stag_w_corr} (red dashed line). %
%
The match to local spin expectation values (inset) is very good.
QMC results for the total magnetization (blue triangles)
match the free spin solution perfectly for fields H/J$_L \le 0.07$.
Above H/J$_L = 0.07$ (corresponding to an applied magnetic field of about 200 T) the magnetization starts to rise significantly, 
which indicates that the applied magnetic field is large enough to break up the singlets.
%thus producing an additional magnetization. 
In simulations of the NMR response (discussed below), the magnetic fields never exceeds H/J$_L$ = 0.01, so that Eq.~\ref{equ:stag_w_corr} remains valid.
%

%
%"""""""""""""""""""""""""""""""""""""""""""""""""""""""""""""""""""""''
\begin{figure}[t]
     \centering
%     \vspace{4mm}
	\includegraphics*[width=0.45\textwidth]{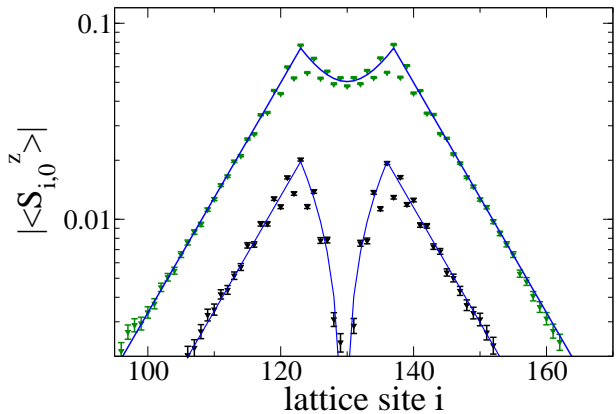}
	\caption{Magnetic moment profiles with two impurities on the same ladder leg. 
        Full lines: description according to Eq.~\ref{equwS02Imp}, symbols: QMC results. 
        Upper curve: impurities are located at positions 123 and 137 on the same sublattice. 
        Lower curve: impurities are located at positions 123 and 136 on different sublattices. 
        QMC simulation for $J_R/J_L=0.5$, $H_{ref}/J_L=0.01$, $T_{ref}/J_L=0.02083$.
    }
	\label{fig:D13and14}      
\end{figure}
%"""""""""""""""""""""""""""""""""""""""""""""""""""""""""""""""""""""''
%
%
%

%-------------------------------------------------------
\subsection{Two impurities: Effective interaction} \label{sec:twoimpurities}
%-------------------------------------------------------

%"""""""""""""""""""""""""""""""""""""""""""""""""""""""""""""""""""""''
\begin{figure}[t]
     \centering
%     \vspace{4mm}
%\includegraphics*[width=0.45\textwidth]{Figures/S2imp_compared_to_2site_model_Screenshot_of_Fig7_of_March.png}
\includegraphics*[width=0.45\textwidth]{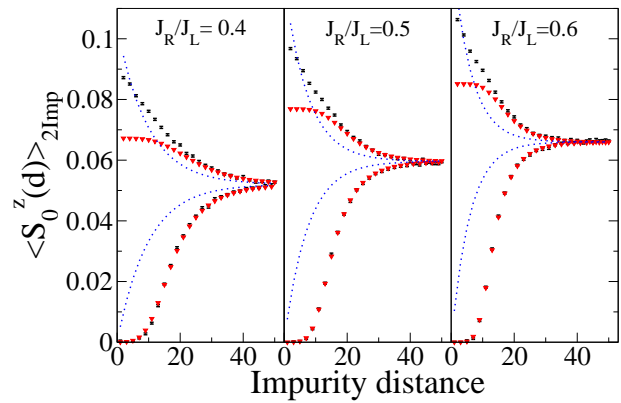}
	\caption{$\langle S^z_0(d)\rangle_{2imp}$ versus distance of two impurities on
         the same ladder leg. QMC results (black error bars) are compared to an effective two-site Heisenberg model (red triangles).
         The blue dotted line corresponds to a simple exponential superposition of two magnetic moment profiles
         calculated from a single impurity.
    ($T/J_L = 0.02083, H/J_L = 0.01$, system size $ 200 \times 2$.)
      }
	\label{fig:S2impAndModel}      
\end{figure}
%"""""""""""""""""""""""""""""""""""""""""""""""""""""""""""""""""""""''
%
%For a realistic simulation, however, one at least has to take the interaction of pairs of close impurities %residing on the same ladder into account. 
The present work is primarily concerned with small impurity concentrations $x\leq0.0025$.
We therefore focus our investigation of impurity interaction on pairs of impurities,
since for combinatoric reasons already three close impurities occur with very small probability.

We find the magnetic moment profile of two impurities to be close to a simple superposition of single-impurity profiles,
but with a modified overall amplitude $\langle S^z_0 \rangle_{2imp}$ instead of $A(\xi_x)$.
%There is, however, an effective interaction between the unpaired spins next to the impurities.
%
At small temperatures where Eq.~\ref{equ_A_of_xi} is valid, 
we find that the T and H dependence 
is again just a \textit{tanh}, and with Eq.~\ref{equ:stag_w_corr} 
the single-impurity profiles to be superposed are 
\begin{equation}
\begin{split}
\langle S^z_{i,j}(T) \rangle_{2imp} = \langle S^z_{0}(H_{ref}/T_{ref},d)\rangle_{2imp} \hspace{2cm}  \\
\quad \times (-1)^{i+j} e^{-\frac{\vert i  \vert}{\xi_x}} \frac{ \tanh (\frac{\mu_B H}{k_B T})}{\tanh (\frac{\mu_B H_{ref}}{k_B T_{ref}})} \quad .
\end{split}
\label{equwS02Imp}
\end{equation}
where $d$ is the distance between the impurities in the x direction, 
\begin{equation}  
 \langle S^z_{0}(\frac{H_{ref}}{T_{ref}},d)\rangle_{2imp}
  =\frac{\langle S^z(i=0,j=0) \rangle_{QMC}}{1+e^{(-\frac{d}{\xi_x})}+e^{(-\frac{N_x-d}{\xi_x})}},
\end{equation}
and $\langle S^z(i=0,j=0)\rangle_{QMC}$ is measured by QMC calculation for 2 impurities. 
The normalization is such that the superposition will reproduce $\langle S^z(i=0,j=0)\rangle_{QMC}$.
Figure ~\ref{fig:D13and14} illustrates the quality of this description by plotting Eq.~\ref{equwS02Imp} 
together with QMC results for two different impurity distances such that the impurities reside on 
(i) the same and (ii) different sublattices of the underlying anti-ferromagnetic structure. 
Changing the impurity distance by one lattice site and thus changing the sublattice leads to very different magnetization profiles. 
This is a consequence of the anti-ferromagnetic order around the impurities, which can be described by two ferromagnetic sublattices that are shifted by one lattice site with respect to each other. If the impurities are an even number of lattice sites apart (same sublattice) their magnetization profiles enhance each other (Fig.~\ref{fig:D13and14}, upper curve). If the impurity distance is an odd number (different sublattice), the profiles interfere destructively leading to  the lower curve in Fig.~\ref{fig:D13and14}. This latter case also resembles the formation of a domain wall.

The unpaired spins interact with each other via the staggered interaction,
which decays exponentially with the spin correlation length.
One may try to describe their interaction with an effective 
two-site Heisenberg model 
\cite{SigristFurusaki96,ImadaIino97,Mikeska97,AnfusoEggert06}
for the two spins located on the same rungs as the impurities,
with coupling $J_{eff}$ and a magnetic field $H$.
We use the ansatz
\begin{equation}\label{eq:EffTwoSpinModel}
   J_{eff} = C(\frac{J_R}{J_L}) \, (-1)^{d-1} \, e^{-(d-1)/\xi_x} ~.
\end{equation}
% C=0.19, 0.245, 0.29 for JR/JL = 0.4, 0.5, 0.6

Figure ~\ref{fig:S2impAndModel} compares the spin expectation values resulting from this
model to QMC results for $\langle S^z_{0}(d)\rangle_{2imp}$ as a function of distance d at different $J_R/J_L$.
We find excellent agreement down to very small distances for the case of odd distances, where the two spins form effective singlets. 
The fitted constants $C(J_R/J_L)$ are 0.19, 0.245, and 0.29 for $J_R/J_L$ = 0.4, 0.5, 0.6, respectively.
For the FM case the agreement is good at large distances, while for small $d$, the two-spin model saturates
whereas the magnetization on the full ladder continues to increase.
Overall, the simple model Eqs.~\ref{equwS02Imp}  and \ref{eq:EffTwoSpinModel}
describes the magnetization data in Figs.~\ref{fig:D13and14}  and \ref{fig:S2impAndModel} 
very well.

%
%"""""""""""""""""""""""""""""""""""""""""""""""""""""""""""""""""""""''
% Stacked ladders, first figure
\begin{figure}[t]
    \centering
%    \vspace{4mm}
	\includegraphics*[width=0.45\textwidth]{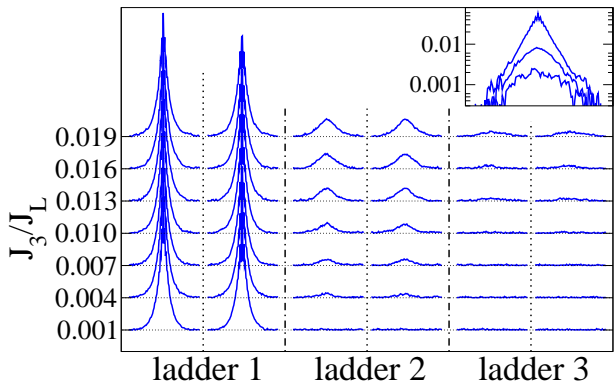}
	\caption{Absolute values of the local spin expectation values  induced by a single impurity ($\corresponds 0.08\dot{3}\%$) for different $J_3$/$J_L$ with $J_R$/$J_L=$ 0.5. The system size is 100 $\times$ 2 $\times$ 6 spins with periodic boundary conditions in leg and stacking direction. $T = 0.02083 J_L \simeq 40$K and $H = 0.01 J_L \simeq 30$ T.
 The impurity resides on the right leg (j=1) of ladder 1 (k=0). For the magnitude of spin expectation values see the inset,
%Imp leg also plotted on the right.
 where the legs with $j=0$ of the same three stacked ladders are plotted at $J_3/J_L=0.019$ on a logarithmic scale,  demonstrating the non-exponential behavior found on neighboring ladders.}
 	\label{fig:Stacked1}
\end{figure}
%"""""""""""""""""""""""""""""""""""""""""""""""""""""""""""""""""""""''
%

%
%%%%%%%%%%%%%%%%%%%%%%%%%%%%%%%%%%%%%%%%%%%%
\section{Stacked Ladders} \label{subsec:stackedlad}
%%%%%%%%%%%%%%%%%%%%%%%%%%%%%%%%%%%%%%%%%%%%
%
%-------------------------------------------------------
%\subsection{Single impurity profile in a stacked system}  \label{sec1ImpStackedDescription}
\subsection{Effective model, single impurity case}  \label{sec1ImpStackedDescription}
%-------------------------------------------------------
%
QMC calculations of up to 8 stacked ladders have been performed using the parameters from Johnston et al. \cite{johnston-2000} In these systems there is an interaction of spins on neighboring ladders via the exchange constant ${J_3}$.
% in addition to the single ladder magnetization profile already described in the previous section. 
%
Spins are now denoted as $S_{i,j,k}$, where $k$ is the distance in stacking direction from the ladder containing the impurity.

Figure ~\ref{fig:Stacked1} shows the spin expectation values on a six-ladder stack with increasing interaction strength J$_3$ and a fixed ratio $J_R$/$J_L=0.5$. We use periodic boundary conditions in stacking direction; 
it is thus sufficient to plot only three ladders. For $J_3$/$J_L \leq$ 0.001 the stacked ladders behave almost like independent single ladders; 
the observed tiny magnetization variations in the neighboring ladder are within the numerical noise. 
With growing $J_3$/$J_L$, an increasing influence on the neighboring ladders is seen. 
Ladder 1 which contains the impurity keeps a simple exponential behavior, with 
a slowly increasing correlation length, from $\xi_x=7.45$ at $J_3=0$ to $\xi_x=9.75$ at $J_3/J_L=0.03$.
However, the effective local magnetic moment distributions on the neighboring ladders do not follow a simple exponential decay law.
Instead, the cusp which appears at $S_0^z$ on the central ladder becomes progressively smeared out on neighboring ladders.

%"""""""""""""""""""""""""""""""""""""""""""""""""""""""""""""""""""""''
\begin{figure}[t]
%	\vspace{-4mm}
   \centering
	\includegraphics*[width=0.45\textwidth]{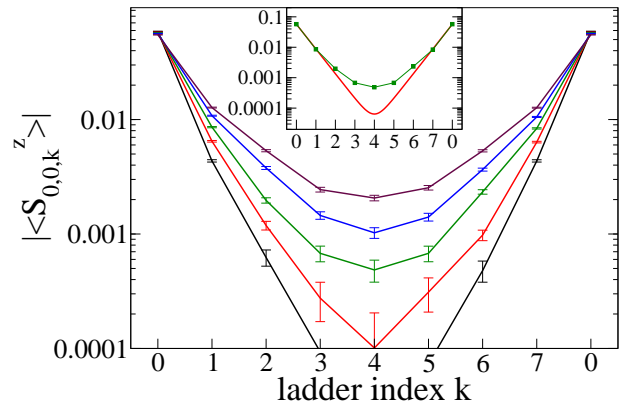}
	\caption{$\vert \langle S_{0,0,k} \rangle \vert$ for a stacked ladder system (k=0,1,2,...,7). $J_R/J_L=0.5$, $H/J_L=0.01$, $T/J_L=0.02083$ and (from bottom to top) $J_3/J_L=0.01,0.015,0.02,0.025,0.03$. The inset compares the QMC results for $J_3/J_L=0.02$ to a \textit{hyperbolic cosine}, demonstrating the non-exponential behavior. Lines are guides to the eye.}
	\label{fig:SmaxLegsDifJ3}      
\end{figure}
%"""""""""""""""""""""""""""""""""""""""""""""""""""""""""""""""""""""''
%
In Fig.~\ref{fig:SmaxLegsDifJ3} the spin expectation values  $\vert \langle S_{0,0,k}^z\rangle \vert$ are plotted for different J$_3$/J$_L$ in stacking direction, from QMC calculations with 8 stacked ladders. 
With increasing J$_3$ the deviation from a simple exponential decay increases. 
This deviation is illustrated in the inset of  Fig.~\ref{fig:SmaxLegsDifJ3} where $\vert \langle S_{0,0,k} \rangle \vert$ is compared with a $\cosh$ for J$_3/J_L=0.02$. We find that for distant ladders the deviation from the \textit{hyperbolic cosine} behavior increases. It thus becomes obvious that a simple exponential cloud model cannot be used to describe the complete profile caused by an impurity in a stacked ladder system.  

Instead we developed a simple effective model, 
related to the one used in Refs.~\onlinecite{Sirker09,SirkerLaflorencie09},
for the actual distribution of magnetizations on stacked ladders.
The ladder containing the impurity (k=0) shows essentially the same profile as a single ladder and can therefore be described by Eq.~\ref{equ:stag_w_corr} 
(with different normalization).

For the other ladders, let us first look at legs with j=0 (opposite the impurity)
We find that the spin expectations value $\langle S^z_{i,0,1} \rangle$ 
on the next-neighboring ladder (k=1) can be calculated by 
treating the spin on each site of the k=0 ladder as a separate source of magnetization,
with an exponentially decaying cloud in leg direction around it on the k=1 ladder.
Subsequently, the magnetizations on the $k=2$ ladder are determined in the same way from those at $k=1$, 
and so on for larger k. 
%considering the influence of all spins on the ladder leg S$^z_{i,0,0}$, which describes the magnetic moment %profile sufficiently well. For the next neighboring ladder with k=2, the profile is calculated by considering %only the influence of S$^z_{i,0,1}$. This procedure is repeated until the end of the stack is reached. 
%Eq.~\ref{equ:feelsstacked} describes the spin profiles on the leg with j=0 for ladders with $k\geq 1$ to be
%
Thus, our ansatz is
\begin{equation}
  \langle S^z_{i,0,k} \rangle = e^{-\frac{1}{\kappa_z}} \sum\limits_{l=-\frac{L_x}{2}}^{\frac{L_x}{2}} 
          (-1)^{\vert l \vert + 1} \langle S^z_{i+l,0,k-1} \rangle e^{-\frac{\vert l \vert}{\xi_x}} ~,
  \label{equ:feelsstacked}
\end{equation} 
where L$_x$ is the length of the ladders
and $\langle S^z_{i,0,0} \rangle$ can be computed like in Eq.~\ref{eq:expdecay}.
This approach treats the magnetization on ladder $k$ like a constant field for the spins on neighboring ladders,
without back-action because of the small value of $J_3$. 
It describes the three-dimensional clouds of magnetizations around impurities very well, as we show below.

The strength of the coupling between the ladders is taken to be a ladder-independent factor $\exp(-1/\kappa_z)$ 
with an effective correlation length $\kappa_z$.
Eq.~\ref{equ:feelsstacked} is to be applied iteratively in k, up to the middle ladder in stacking direction,
with the remaining ones determined by symmetry
%Exploiting the profile symmetry for the stacked ladders around the ladder containing the impurity 
\begin{equation}
\langle S^z_{i,j,-k} \rangle = \langle S^z_{i,j,k} \rangle
\end{equation}

For the other ladder legs (j=1), we approximate the spin expectation values as 
\begin{equation}
\langle S^z_{i,1,k} \rangle = -\langle S^z_{i,0,k} \rangle 
\end{equation}
for $k\ge1$.

To determine $\kappa_z$ we apply Eqs. \ref{equ:stag_w_corr} %, \ref{equ_A_of_xi}, 
and \ref{equ:feelsstacked}                                  %for H=H$_{ref}$ and T=T$_{ref}$ and obtains
to get 
\begin{equation}
  \langle S^z_{0,0,1} \rangle = - \langle S^z_{0,0,0} \rangle e^{-\frac{1}{\kappa_z}} \left( 1+2 \sum\limits_{l=1}^{\frac{L_x}{2}} e^{-\frac{2 l}{\xi_x}} \right)
  \label{equ:extractKappa}
\end{equation}
with $\langle S^z_{0,0,1} \rangle$, $\langle S^z_{0,0,0} \rangle$ and $\xi_x$ 
taken from a QMC simulation with one impurity
on stacked ladders. 
For the interladder couplings $J_3/J_L$ between 0.01 and 0.03 studied in Sec. \ref{sec:Results}, we find $\kappa_Z$ to be small, ranging from 0.22 to 0.28.
The magnetization profile, Eq.~\ref{equ:feelsstacked}, is thus computed from 
only these three measured quantities. 
%
%
%"""""""""""""""""""""""""""""""""""""""""""""""""""""""""""""""""""""''
\begin{figure}[t]
  \centering
%	\vspace{7mm}
	\includegraphics*[width=0.45\textwidth]{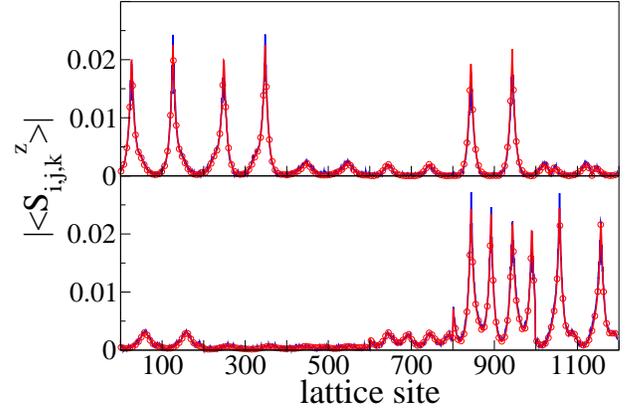}
	\caption{Local spin expectation values (absolute values) for 6 stacked ladders with 3 impurities.  Spin sites 1 - 200 correspond to ladder 1, 201 - 400 to ladder 2, and so on. The red line (open circles) is the result of our model function, the underlying blue solid line is the corresponding QMC simulation. Upper panel: impurities located on 3 different ladders [(27,0,0),(50,0,1),(46,0,4)]. Lower panel: 2 impurities located on the same ladder [(44,1,4),(92,1,4)] and one on a different ladder [(59,1,5)]}
	\label{fig:SpinDist}
\end{figure}
%"""""""""""""""""""""""""""""""""""""""""""""""""""""""""""""""""""""''

%-------------------------------------------------------
\subsection{Multiple impurities}
%-------------------------------------------------------

Similar to single ladders, multiple impurities on stacked ladders can be described by superimposing single-impurity profiles, Eq.~\ref{equ:feelsstacked}.
When the impurities are located on the same ladder, we account for their interaction by
using $\langle S^z_0(d) \rangle_{2imp}$, 
Eq.~\ref{equwS02Imp}, measured on the stacked ladder system,
in place of $\langle S^z_{0,0,0} \rangle$.
Since $J_3$ is very small, two impurities located on different ladders do not influence each other much,
and we use the single-ladder results in this case.
Larger numbers of close-by impurities are extremely rare at small concentrations $x$,
so that we can omit the effect of their coupling on the NMR spectrum.

In Fig.~\ref{fig:SpinDist}, we compare our model to QMC results for two distinctly different impurity distributions, 
namely 3 impurities distributed over 3 ladders (upper panel),
% of Fig.~\ref{fig:SpinDist}), 
and 2 impurities on the same ladder and one on an adjacent ladder (lower panel). % of Fig.~\ref{fig:SpinDist}).
We find excellent agreement.
Let us emphasize that even such complex spin profiles require only very few parameters for the model function, 
calculated at a reference  temperature and reference external magnetic field,
namely $\xi_x$, $\langle S^z_0 \rangle$, $\langle S^z_{0,0,1} \rangle$,
and $\langle S^z_0(d) \rangle_{2imp}$.
Only the last of these depends on impurity positions.

%
%%%%%%%%%%%%%%%%%%%%%%%%%%%%%%%%%%%%%%%%%%%%
\section{NMR } \label{sec:NMR-Simulation}
%%%%%%%%%%%%%%%%%%%%%%%%%%%%%%%%%%%%%%%%%%%%
%
%-------------------------------------------------------
\subsection{Magnetic moment profile and calculation of the NMR spectrum} \label{secProfile2NMR}
%-------------------------------------------------------
%

%"""""""""""""""""""""""""""""""""""""""""""""""""""""""""""""""""""""''
\begin{figure}[t]
	\vspace*{-0mm}
   \centering
	\includegraphics*[width=0.48\textwidth]{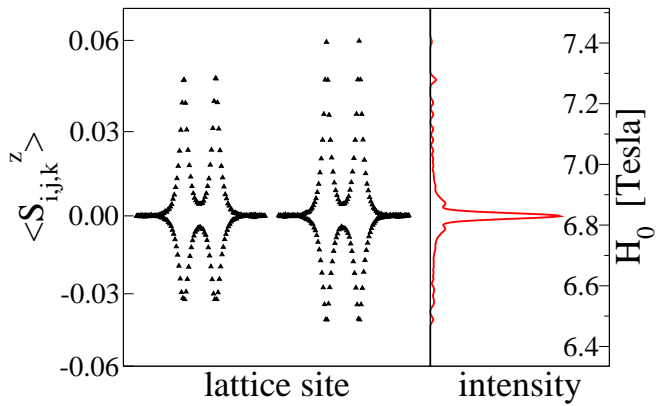}
	\caption{Relation between local magnetic moment profiles and the corresponding NMR spectrum.
     Left panel: Spin expectation values (left and right leg)  of a ladder containing two impurities 
         on the left leg.
      Right panel: The corresponding NMR spectrum reflects the histogram of spin expectation values.
        The experimentally observed broadening corresponds to small values of $\langle S^z_{i,j,k} \rangle$
         up to about 0.01.}
	\label{fig:Profile2Spec}      
\end{figure}
%"""""""""""""""""""""""""""""""""""""""""""""""""""""""""""""""""""""''
%
%"""""""""""""""""""""""""""""""""""""""""""""""""""""""""""""""""""""''
\begin{figure}[t]
%	\vspace{5mm}
   \centering
	\includegraphics*[width=0.45\textwidth]{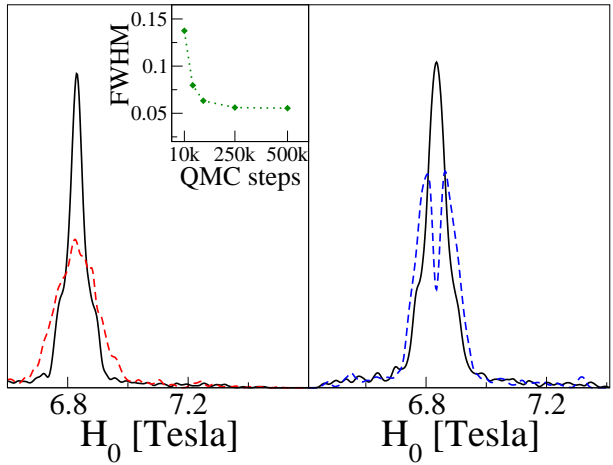}
	\caption{NMR spectra with three impurities at fixed positions. 
     Left panel: Influence of the stochastic noise on the width. 
        Impurities at positions (7,1,2), (51, 2, 6), (73, 1, 5), 
        with QMC simulations of  10k (red dashed line) and 500k (black solid line) sweeps.
        The inset shows the dependence of the FWHM on the number of steps. 
     Right panel: Two NMR spectra of two systems only differing in impurity configuration (250k steps).
       Black solid line: same as left panel; blue dashed line: positions (35,1,1), (52,1,4), (86,2,5).
     (J$_R$/J$_L$=0.5, J$_3$/J$_L$=0.017, T/J$_L$=0.02083, H/J$_L$=0.003828.)
}
	\label{figQMCprobNMR}      
\end{figure}
%"""""""""""""""""""""""""""""""""""""""""""""""""""""""""""""""""""""''

To obtain the Cu$^{2+}$ NMR spectrum influenced by a magnetic moment profile one has to consider the spin-ion hyperfine coupling in the resonance condition\cite{PhysRevLett.83.412}
\begin{equation}
  \frac{\nu_{RF}}{\gamma_{N}}=H_{0}(i,j,k) + A_{HF} \mu_B g \langle S_{i,j,k}^{z}(H_0(i,j,k))\rangle \quad.
  \label{equResCon}
\end{equation}
$H_0(i,j,k)$ is the value of the external field $H_0$ matching the resonance condition at site $i,j,k$. Here, $\nu_{RF}$ is the frequency of the RF field, $\gamma_{N}$ is the nuclear gyro-magnetic ratio, $A_{HF}$ is the hyperfine
coupling (-12T/$\mu_B$ for SrCu$_2$O$_3$ \cite{PhysRevLett.80.604}), 
and $\langle S_{i,j,k}^{z} (H_0(i,j,k))\rangle$ is the expectation value for the $z$ component of a spin at site $i,j,k$ induced by an external field $H_0$.

The dependence on temperature and magnetic field of the spin expectation value is the {\em tanh} discussed before, so that QMC calculations need only be done at some reference temperature $T_{ref}$ and reference field $H_{ref}$:
\begin{equation}
\begin{split}
\frac{\nu_{RF}}{\gamma_{N}}=H_{0}(i,j,k) \hspace{45mm}\\
  + A_{HF} \mu_B g \langle S^z_{i,j,k}(H_{ref}) \rangle \frac{tanh(\frac{\mu_B H_0(i,j,k)}{k_B T})}{tanh(\frac{\mu_B H_{ref}}{k_B T_{ref}})}
  \label{equCompleteResCon}
\end{split}
\end{equation}
The NMR spectrum is then obtained by collecting the values of $H_0(i,j,k)$  from all lattice sites $(i,j,k)$ into a histogram. 
To account for the natural linewidth we convolute the histograms with a Gaussian of 0.02T half linewidth. %

Figure ~\ref{fig:Profile2Spec} illustrates the resulting NMR spectrum arising from two impurities on a single ladder,
with magnetic field strengths as used in the experiment by Fujiwara et al. (Fig.~\ref{fig:ExpFuji}).
Note that the broadening seen in experiment corresponds to very small values of $\langle S^z_{i,j,k} \rangle$,
up to about 0.01.
Larger spin expectation values, which occur on only few sites close to impurities, 
contribute only little to the histogram, in its outliers.
For the understanding of the NMR signal such large spin expectation values may therefore be
treated more approximately, as we do in our model descriptions.

Figure~\ref{figQMCprobNMR} illustrates two effects which influence the calculated distribution.
The left panel shows that stochastic noise of small spin expectation values in the QMC calculation 
produces an effective broadening.
Therefore QMC calculations of high precision are necessary in order to get  reliable NMR histograms.
In the right panel of Fig.~\ref{figQMCprobNMR} we show the influence of different impurity positions 
on the linewidth and line shape, which demonstrates the necessity to average over a large number of impurity configurations.

If each impurity configuration needed to be simulated in a separate QMC simulation, the computational
effort would be too large.
Instead, we employ the effective analytical description given in Sec. \ref{sec1ImpStackedDescription},
which allows us to calculate the NMR spectra for many impurity configurations 
on the basis of only a few parameters measured in QMC simulations.
For each set of couplings $J_R/J_L$ and $J_3/J_L$ 
we calculated $\xi_x$, $\langle S^z_0 \rangle$,
and $\langle S^z_{0,0,1}\rangle$ on a $200\times 2\times 6$ system of coupled ladders
with one impurity ($200\times 2\times 8$ for $J_3/J_L=0.03)$.
We calculated $\langle S^z_0 \rangle_{1imp}$ and $\langle S^z(d) \rangle_{2imp}$  for distances $d=1\dots 40$ of two impurities on the same leg of a ladder at $J_3/J_L=0.01$ on $200\times 2\times 4$ coupled ladders.
Since this calculation was very time consuming, 
we used the ratio between $\langle S^z(d) \rangle_{2imp}$ and $\langle S^z_0 \rangle_{1imp}$ also at other values of $J_3$.
% N.B. (it should barely depend on $J_3$)
All QMC calculations for the NMR spectra were done at the reference values $T_{ref}= 0.02083 J_L = 40 K$ and $H_{ref}=0.01 J_L\simeq 19$ T.
%and where extended to other values with Eq.~\ref{equCompleteResCon}.

%"""""""""""""""""""""""""""""""""""""""""""""""""""""""""""""""""""""''
% First figure of results Sec.
\begin{figure}[t]
   \centering
	\includegraphics*[width=0.40\textwidth]{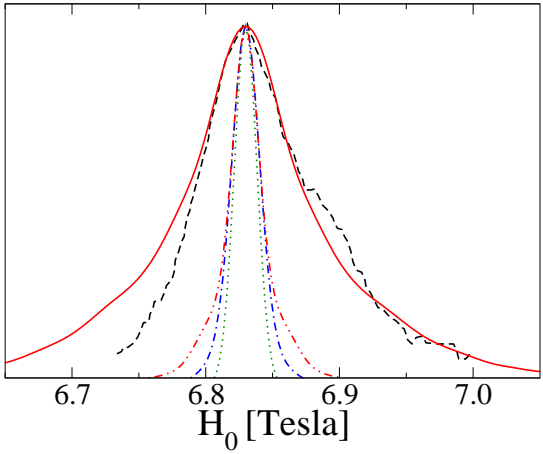}
	\caption{Simulated NMR spectra for $x=0.25\%$ impurities at T = 40 K, $J_L=$ 1920 K, and $J_R$/$J_L=$ 0.5 
               with stacked ladder couplings $J_3$/$J_L=$ 0.01 (red dot-dot-dashed line) 
               and $J_3$/$J_L=$ 0.03 (solid red line), 
               compared to experiment (Fig.~\ref{fig:ExpFuji}, black dashed line). 
               Sizeable broadening occurs only for the larger interladder coupling.
               At high temperature (340 K), all spectra are narrow.
               For reference, we show an undoped system (green dotted line, Gaussian line shape) 
               and the doped system (blue dot-dashed line) at $J_3/J_L=0.03$. 
            }
	\label{fig:NMRCompareJohnston}
\end{figure} 
%"""""""""""""""""""""""""""""""""""""""""""""""""""""""""""""""""""""''

%-------------------------------------------------------
\subsection{Results}\label{sec:Results}
%-------------------------------------------------------
%

%"""""""""""""""""""""""""""""""""""""""""""""""""""""""""""""""""""""''
\begin{figure}[t]
%	\vspace{4mm}
    \centering
	\includegraphics*[width=0.45\textwidth]{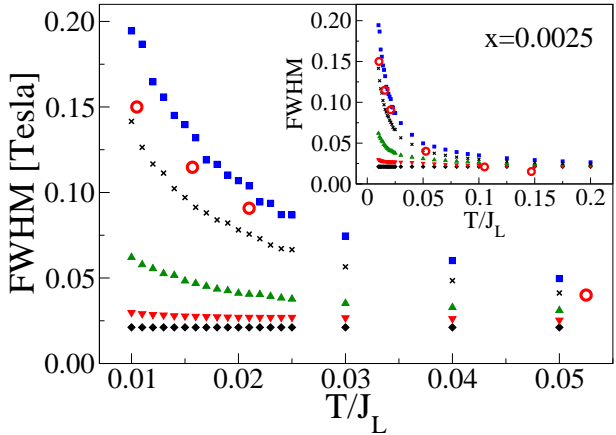}
	\caption{FWHM vs.\  temperature for a stacked ladder system with J$_L=$ 1920K, $J_R$/$J_L=$ 0.5, and varying inter ladder coupling $J_3$/$J_L$: 0.0 (black diamonds), 0.01 (red down-triangles), 0.02 (green up-triangles), and 0.03 (blue squares) . The black crosses show results
with a different rung coupling $J_R/J_L=0.4$ at $J_3/J_L=0.02$.
% The individual correlation lengths for each curve are extracted from a QMC calculation around $T/J_L=$ 0.02083 (=40 K) and assumed to be independent of temperature. 
Experimental values 
for the $^{65}$Cu central peak
from Fig.~\ref{fig:ExpFuji} 
are given by red open circles. 
The inset shows an expanded temperature range.}
\label{fig:FWHMvsTemp}
\end{figure}
%"""""""""""""""""""""""""""""""""""""""""""""""""""""""""""""""""""""''
%
%"""""""""""""""""""""""""""""""""""""""""""""""""""""""""""""""""""""''
\begin{figure}[t]
%   \vspace{3mm}
   \centering
	\includegraphics*[width=0.45\textwidth]{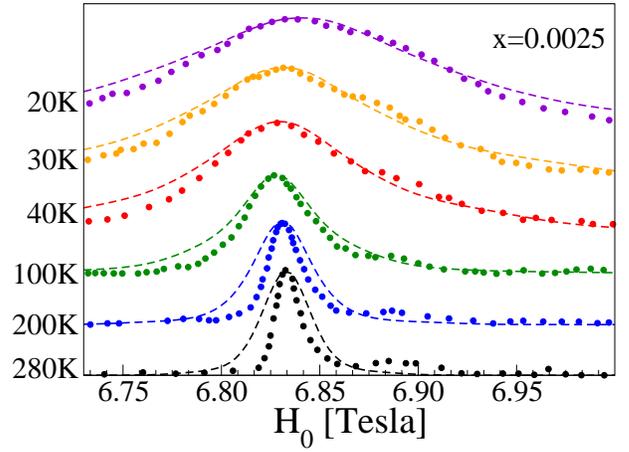}
	\caption{Simulated NMR spectra at $x=0.0025$, with  J$_L$= 1920K, $J_R$/$J_L=$ 0.5, and $J_3$/$J_L=$ 0.03 and different temperatures compared to experiment\cite{PhysRevLett.80.604} 
($^{65}$Cu left peak, $\nu_{RF}=83.55MHz$, filled circles).}
\label{fig:NMRSimVsFujiwara}
\end{figure}
%"""""""""""""""""""""""""""""""""""""""""""""""""""""""""""""""""""""''
%"""""""""""""""""""""""""""""""""""""""""""""""""""""""""""""""""""""''
\begin{figure}[t]
%   \vspace{4mm}
   \centering
	\includegraphics*[width=0.45\textwidth]{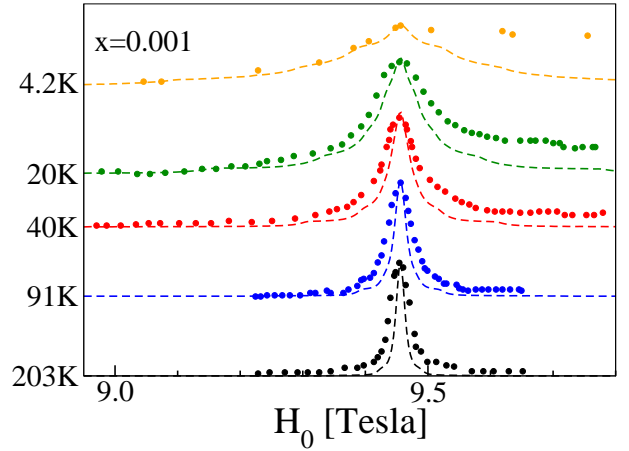}
	\caption{
Simulated NMR spectra at $x=0.001$, with J$_L$= 1920K $J_R$/$J_L=$ 0.5, and $J_3$/$J_L=$ 0.03 and different temperatures compared to experiment\cite{PhysRevB.60.4181}
($^{65}$Cu left peak, $\nu_{RF}=125.1MHz$, filled circles).
The asymmetric experimental profiles below 40 K are caused by an overlap of a $^{63}$Cu transition with its main peak around 10.05 T.
The experiment appears to have a larger natural linewidth than assumed in our simulations.}
\label{fig:NMRCompareOhsugi0o1perc}
\end{figure}
%"""""""""""""""""""""""""""""""""""""""""""""""""""""""""""""""""""""''
%
%"""""""""""""""""""""""""""""""""""""""""""""""""""""""""""""""""""""''
\begin{figure}[t]
%   \vspace{4mm}
   \centering
	\includegraphics*[width=0.49\textwidth]{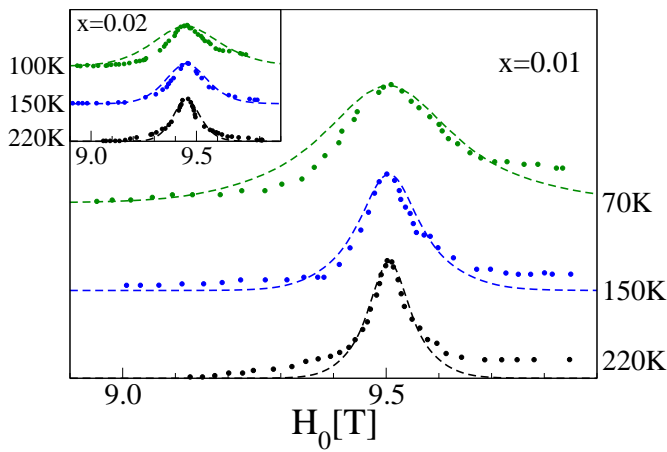}
	\caption{
Simulated NMR spectra at $x=0.01$, with J$_L$= 1920K $J_R$/$J_L=$ 0.5 and $J_3$/$J_L=$ 0.03 and different temperatures compared with experiment\cite{PhysRevB.60.4181}
($^{65}$Cu left peak, filled circles) 
The inset shows the same system with an impurity concentration of x=0.02.}
	\label{fig:NMRCompareOhsugi1perc}
\end{figure}
%"""""""""""""""""""""""""""""""""""""""""""""""""""""""""""""""""""""''
%

We used our effective model to generate NMR spectra for given impurity concentrations
and for different couplings, based on the QMC simulations described above.
10$^4$ random vacancy configurations were generated 
on systems of up to $400\times 2\times 20$ sites
for each set of couplings $J_R/J_L$ and $J_3/J_L$.
The NMR spectrum was calculated for each vacancy configuration using our model,
and the results were superimposed to obtain spectra comparable to realistic NMR signals. 

In Fig.~\ref{fig:NMRCompareJohnston} we show results at $x=0.25\%$ impurity doping 
for two different stacked ladder couplings J$_3$/J$_L$=0.01 and J$_3$/J$_L$=0.03.
%J$_R$/J$_L$=0.5, J$_L$=1920K, 
%
At high temperature (340 K), all spectra are similar to the pure Gaussian line shape,
i.e.\ no broadening is visible, even at the larger ladder coupling $J_3/J_L=0.03$.
This is due to the suppression of magnetic moments by the factor $\tanh (\frac{\mu_B H}{k_B T})$.
At low temperature, T=40 K, the NMR spectrum remains narrow for isolated ladders (not shown)
since an exponential decay with the small correlation length of the undoped system
(Fig.~\ref{fig:LadderLegsOverLay}) does not contain enough sites 
with the relevant range of magnetizations.
The spectrum remains narrow also for small interladder coupling $J_3/J_L=0.01$.

The behavior changes drastically for the larger interladder coupling $J_3/J_L=0.03$.
Then several stacked ladders obtain magnetizations within the relevant range
(cf.\ Fig.~\ref{fig:SmaxLegsDifJ3}), resulting in a broadened NMR spectrum
in excellent agreement with the experimental spectrum.

In Fig.~\ref{fig:FWHMvsTemp} we study the broadening at $x=0.25\%$ in more detail.
We show the FWHM (\textbf{F}ull \textbf{W}idth at \textbf{H}alf \textbf{M}aximum) versus temperature for different interladder couplings $J_3$.
The couplings are J$_R$/J$_L$=0.5 and
J$_3$/J$_L$=0.0 (resulting in $\xi_x$=7.45),  
0.01 ($\xi_x$=7.6, $\kappa_z$=0.218),  
0.02 ($\xi_x$=8.6,  $\kappa_z$=0.253),  
0.03 ($\xi_x$=9.75, $\kappa_z$=0.278),
as well as J$_R$/J$_L$=0.4 with J$_3$/J$_L$=0.02 ($\xi_x$=11.3554, $\kappa_z$=0.254). 
We find that simulated spectra with a stacked ladder coupling slightly below  $0.03 J_L$,
at the upper end of the couplings specified in Ref.~\cite{johnston-2000},
match the experimental results well over the whole temperature range.
The correlation lengths and $\kappa_z$ were extracted at $T=0.02083 J_L = 40 K$ and assumed to be temperature-independent.
%to demonstrate that the broadening effect is primarily due to the changes in the spin distributions. 
This assumption is valid for the low-dilution, low-temperature regime, 
while for higher temperature there is little broadening so that the influence of the correlation length becomes unimportant. 

In Figs.~\ref{fig:NMRSimVsFujiwara},\ref{fig:NMRCompareOhsugi0o1perc}, and \ref{fig:NMRCompareOhsugi1perc}, 
we show a detailed comparison of our simulated NMR spectra with experimental results
at impurity concentrations of $0.25$, $0.1\%$ (which appears to have a larger natural linewidth),
and $1\%$, using J$_R$/J$_L$=0.5 and J$_3$/J$_L$=0.03.
%(resulting in $\xi_x$=9.75 , $\kappa_z$=0.278 , $\langle S_0^z \rangle_{1imp}=0.0475$ at T/J$_L$=0.02083
%
%x=0.0025 (Fig.~\ref{fig:NMRSimVsFujiwara} and Ref.~\onlinecite{PhysRevLett.80.604}), 
%x=0.001  (Fig.~\ref{fig:NMRCompareOhsugi0o1perc} and Ref.~\onlinecite{PhysRevB.60.4181}), 
%and x=0.01 (Fig.~\ref{fig:NMRCompareOhsugi1perc}, Ref.~\onlinecite{PhysRevB.60.4181}).
We find that our model is in very good agreement with experiment in almost all cases.
%

%%%%%%%%%%%%%%%%%%%%%%%%%%%%%%%%%%%%%%%%%%%%
\section{Conclusions} \label{sec:conclusions}
%%%%%%%%%%%%%%%%%%%%%%%%%%%%%%%%%%%%%%%%%%%%
We studied impurity-containing systems of stacked spin ladders by means of QMC simulations. 
Temperature and field dependence of magnetizations are paramagnetic.
We developed an effective spin model for the interaction of unpaired spins next to two impurities.
In contrast to assumptions made in earlier investigations 
%\cite{Bobroff09,PhysRevB.81.054438} 
we observed that the staggered magnetization caused 
by an impurity does not follow a simple  three-dimensional exponential behavior. The spin distributions on the ladders in stacking direction deviate progressively from such an exponential dependence.  
We provided an analytical description for the spin profiles
in systems with multiple impurities and used it to simulate the NMR spectra of lightly doped SrCu$_2$O$_3$
with only a small number of parameters determined by QMC.
The resulting NMR spectra allowed us to explain the drastic broadening of the $^{65}$Cu NMR line in SrCu$_2$O$_3$ found in experiments\cite{PhysRevLett.80.604, PhysRevB.60.4181} at intermediate temperatures
to be a consequence of a sizable coupling between ladders in stacking direction.
which causes a non-exponential cloud of small effective magnetic moments to occur around impurities.

%%%%%%%%%%%%%%%%%%%%%%%%%%%%%%%%%%%%%%%%%%%%
\begin{acknowledgments}
%%%%%%%%%%%%%%%%%%%%%%%%%%%%%%%%%%%%%%%%%%%%
The authors acknowledge support from the Austrian Science Fund FWF within  
SFB ViCoM F4109-N13 P04 and P09 (H.G.E. and P.M.) and from the Science College W401-N13 (R.A.).
\end{acknowledgments}

\bibliography{literature}

\end{document}